\documentstyle[amssymb,pre,twocolumn,aps]{revtex}

\begin{document}
\title{Topological Defects and Non-homogeneous Melting of Large 2D Coulomb Clusters }
\author{Minghui Kong, B. Partoens, and F. M. Peeters\cite{pet}}
\address{{\it Departement Natuurkunde, Universiteit Antwerpen (UIA)}\\
{\it Universiteitsplein 1, B-2610 Antwerpen, Belgium}}
\maketitle

\begin{abstract}
The configurational and melting properties of large two-dimensional (2D)
clusters of charged classical particles interacting with each other via the
Coulomb potential are investigated through the Monte Carlo simulation
technique. The particles are confined by a harmonic potential. For a large
number of particles in the cluster (N%
\mbox{$>$}%
150) the configuration is determined by two competing effects, namely in the
center a hexagonal lattice is formed, which is the groundstate for an
infinite 2D system, and the confinement which imposes its circular symmetry
on the outer edge. As a result a hexagonal Wigner lattice is formed in the
central area while at the border of the cluster the particles are arranged
in rings. In the transition region defects appear as dislocations and
disclinations at the six corners of the hexagonal-shaped inner domain. Many
different arrangements and type of defects are possible as metastable
configurations with a slightly higher energy. The particles motion is found
to be strongly related to the topological structure. Our results clearly
show that the melting of the clusters starts near the geometry induced
defects, and that three different melting temperatures can be defined
corresponding to the melting of different regions in the cluster.

\end{abstract}

\section{INTRODUCTION}

Recently, there has been considerable theoretical and experimental progress
in the study of mesoscopic systems consisting of a finite number of charged
particles which are confined into an artificial circular symmetric
potential. In 1934, Wigner suggested that a liquid to solid phase transition
should occur in a three-dimensional (3D) Fermi system at low densities \cite
{1}. Typical experimental model systems for the study of this system are
electrons on the surface of liquid helium \cite{2}, electrons in quantum
dots \cite{nature}, colloidal suspensions\cite{2.5} and in confined plasma
crystals\cite{2.6}. On the other hand, various similar systems, like the
vortex clusters in an isotropic superfluid \cite{15}, vortices in superfluid
He$^{4}$\cite{1617}, vortices in a Bose-Einstein condensate stirred with a
laser beam \cite{18} and in mesoscopic superconducting disks \cite{19} have
many common features with those of 2D charged particles. Colloidal particles
dissolved in water \cite{20.1} are another example of an experimental system
where classical particles exhibit Wigner crystallization. Recently,
macroscopic 2D Wigner islands, consisting of charged metallic balls above a
plane conductor were studied and ground state, metastable states and saddle
point configurations were found experimentally \cite{24}.

Such a system with a finite number of particles, initially studied by
Thomson as a classical model for the atom \cite{Thomson,Thomson2}, has been
extensively studied during the past few years. For a small number of
particles (typically $N<100$) they are arranged in rings \cite{27.5,25,26,27}
and a Mendeleev-type of table was constructed in Ref. \cite{25} which gives
the distribution of those particles over the different rings. Moreover, the
configurations of the ground state, the metastable states and saddle point
states were obtained, from which the transition path and the geometric
properties of the energy landscape were given in Ref. \cite{kong}. The
spectral properties of the ground state configurations were presented in
Ref. \cite{26} and generalized\ to screened Coulomb \cite{31,30} and
logarithmic \cite{Thomson2,30} interparticle interactions. The excitation of
normal modes of 2D Coulomb clusters in laboratory complex plasmas were
recently observed \cite{Melzer}.

The melting properties of this system have been studied experimentally \cite
{20.1,Zahn} and by Monte Carlo(MC) \cite{SchPRL82(1999),Loz2001PRL} and
molecular dynamics \cite{Sch2000,LinI2001} simulations. In a hard wall
confined system with short-range inter-particle interaction, the melting
behavior was found even more interesting. Reentrant melting of 2D colloidal
clusters in a hard wall potential was obtained in both experimental \cite
{20.1} and theoretical work \cite{Sch2000}.

The defect structure in crystals is of paramount importance for the
stability and the strength of materials. Topological defects in Wigner
crystals~\cite{Ceperley,xsling} and its effect on particle melting were
investigated in Refs.~\cite{Zahn,SchPRL82(1999),LinI2001,Moore}. Thermal
defect mediated melting was proposed as the microscopic mechanism for
melting in an infinite 2D triangular Wigner crystal. It is well known that
the KTHNY scenario(Kosterlitz-Thouless-Halperin-Nelson-Young) describes 2D
melting as a defect-mediated phenomenon and melting occurs in two stages
through two continuous phase transitions \cite{Peeters}. A clear case of
two-stage melting was observed in a film of paramagnetic colloidal particles 
\cite{Zahn}.

In this paper we study topological defects which are induced by the
confinement potential, i.e. which are a result of the finite size of the
system. Next we investigate how these defects influence the melting of the
mesoscopic 2D island. The present paper is organized as follows. In Sec. II,
we describe the model system and the numerical approach. Sec. III is devoted
to the structural properties of the topological defects at zero temperature.
In Sec. IV, we discuss the eigenmode spectrum for these large clusters. The
discussion on the non-homogeneous melting is presented in Sec. V. Our
conclusions are given in Sec. VI.

\section{Numerical Approach}

The model system was defined in Ref. \cite{25} and the Hamiltonian for such
a system is given by 
\begin{equation}
H=\frac{q^{2}}{\varepsilon }\sum_{i>j}^{N}\frac{1}{\left| \stackrel{%
\rightarrow }{r_{i}}-\stackrel{\rightarrow }{r_{j}}\right| }+\sum_{i}^{N}V(%
\stackrel{\rightarrow }{r_{i}}).
\end{equation}
The confinement potential $V(\stackrel{\rightarrow }{r})=\frac{1}{2}m^{\ast
}\omega _{0}^{2}r^{2}$ is taken circular symmetric and parabolic, where $%
m^{\ast }$ is the effective mass of the particles, $q$ is the particle
charge, $\omega _{0}$ is the radial confinement frequency and $\varepsilon $
is the dielectric constant of the medium the particles are moving in. To
exhibit the scaling of the system, we introduce the characteristic scales in
the problem: $r_{0}=(2q^{2}/m\epsilon \omega _{0}^{2})^{1/3}$ for the
length, $E_{0}=(m\omega _{0}^{2}q^{4}/2\epsilon ^{2})^{1/3}$ for the energy
and $T_{0}=(m\omega _{0}^{2}q^{4}/2\epsilon ^{2})^{1/3}k_{B}^{-1}$ for
temperature. After the scaling transformations ($r\rightarrow
r/r_{0},E\rightarrow E/E_{0},T\rightarrow T/T_{0}$), the Hamiltonian can be
rewritten in a simple dimensionless form as

\begin{equation}
H=\sum_{i>j}^{N}\frac{1}{\left| \stackrel{\rightarrow }{r_{i}}-\stackrel{%
\rightarrow }{r_{j}}\right| }+\sum_{i}^{N}r_{i}^{2},
\end{equation}
which only depends on the number of particles $N$. The numerical values for
the parameters $\omega _{0},r_{0},E_{0},T_{0\text{ }}$for some typical
experimental systems were given in Ref. \cite{25}.

The MC simulation technique \cite{36} is relatively simple and rapidly
convergent and it provides a reliable estimation of the total energy of the
system in cases when relatively small number of Metropolis steps is
sufficient. However, the accuracy of this method in calculating the explicit
states is poor for systems with a large number of particles, which have
significantly more metastable states. To circumvent this problem we employed
the Newton optimization technique which was outlined and compared with the
standard Monte Carlo technique in Ref. \cite{26}. The structure and
potential energy of the system at $T$ $\neq 0$ are found by the standard
Metropolis algorithm \cite{36} in which at some temperature the next
simulation state of the system is obtained by a random displacement of one
of the particles. We allow the system to approach its equilibrium state at
some temperature $T$, after executing $10^{4}-5\times 10^{5}$ ``MC steps''.
Each MC step is formed by a random displacement of all particles. If the new
configuration has a smaller energy it is accepted, but if the new energy is
larger the configuration is accepted with probability $\delta <\exp (-\Delta
E/T),$ where $\delta $ is a random number between 0 and 1 and $\Delta E$ is
the increment in the energy \cite{SchPRL82(1999)}.

\section{Topological defects}

It is well known that the hexagonal lattice is the most energetically
favored structure for classical point charges in a two-dimensional infinite
plane at low temperature \cite{79}. For a system consisting of a finite
number of repelling particles restricted to 2D, which are held together by a
circular harmonic potential, the cluster patterns are determined by the need
to balance the tendency to form a triangular lattice against the formation
of a compact circular shape. The configuration is determined by these two
competing effects, namely circular symmetry and triangular structure (Wigner
lattice). This competition leads to intrinsic defects in the 2D circular
Coulomb cluster which are {\it geometry} (of the confinement potential) {\it %
induced defects}. This ground state is not a defect free system. The
symmetry breaking is due to the packing of the triangular lattice into a
region with a circular boundary. A hexagonal lattice which is cut by a
circle without the introduction of any defect has an energy $E=56.0499E_{0}$
which is larger than the ground state energy $E=55.9044E_{0}$ for $N=291$
particles.

In the first part of this paper, we investigate the form and position of the
defects in large clusters. Therefore we make use of the Voronoi construction 
\cite{voronoi}. The Voronoi construction of a collection of particles
consists of a partition of space into cells. Each cell consists of those
points which are closer to the given particular particle than to any other
particles. Examples of Voronoi constructions are shown in Fig.~\ref{Fig1}
where the ground state configuration for $N=291,300,400$ and $500$ are
shown. One can see that there are two kinds of defects, i.e. dislocations
and disclinations. Disclinations are orientational defects with five
(indicated by `--') or seven (indicated by `+') fold coordination number
(the number of sides of the polygon around the particles is nothing else
then the coordination number). A dislocation is a pair of two disclinations
consisting of a defect with 5-fold (--) and a defect with 7-fold (+)
coordination number. In the latter case the ordering at long distances is
not disrupted and consequently such a bound pair has a much lower energy 
\cite{Peeters}. The total number of 5-fold $N_{-}$ and 7-fold $N_{+}$
disclinations depends on the particular configuration. The number of
disclinations in this system is determined by Euler's theorem and can't be
changed, so the net topological charge $N_{-}-N_{+}$ is always equal to six
as was already demonstrated in Refs. \cite{30,koulakov}. The reason is that
every `--' defect can bend the lattice clockwise over $\pi /3$ from a
straight lattice and thus six `--' defects can bend a straight line into a
circle. Dislocations will appear when it decreases the energy of the system.
From Fig. 1 it is apparent that this is more so for larger clusters.

In Refs.~\cite{30,LinI2001}, the defects in clusters with a logarithmic
inter-particle interaction were studied. We want to stress that their way of
visualizing the defects is different: nearest neighbours are connected by a
line, without making crossings. However, this does not lead to a unique
picture: the total number of 5-fold, $N_{-},$ and 7-fold, $N_{+},$
disclinations can vary in the same configuration, only the net topological
charge $N_{-}-N_{+}$ is always equal to six.

In these large clusters, the defects are located on a hexagon, i.e. they
form a hexagonal structure. As can be seen in Fig.~\ref{Fig1} the defects
are approximately situated at the six corners of a hexagon, each corner with
a net topological charge of `--1'. Notice from Fig.~\ref{Fig1} that a single
5-fold disclination can appear, but never a single 7-fold disclination. For
the large strain energy around `--1' topological charge, some dipole
defects(i.e. dislocation with `--1' and `+1' defects) will be generated to
shield the `--1' topological charge. Those shielding dipole defects do not
change the topology of the system. A clearer example is shown in Fig.~\ref
{Fig2} for the $N=291$ groundstate configuration. We considered the $N=291$
system as it minimizes the number of defects. The reason is that for this
particle number the configuration has 42 particles in the outer ring, which
is a multiple of the topological charge. There are three rings at the border
with an equal number ($N=42$) of particles (the 1D Wigner lattice), the
central hexagonal structure (the 2D Wigner lattice) and the defects
indicated by triangles ($\bigtriangleup $) and squares ($\square $) are
situated: (i) around the six corners of a hexagon and (ii) in the transition
region between the outer rings and the central hexagon.

It should be noted that the search for the global minimum configuration is a
difficult problem for large systems because of the existence of a large
number of local-minimum configurations, with energy very close to the global
minimum. Thus one is never $100\%$ sure to have found the real ground state.
Therefore, we investigated the different metastable states. In an
experiment, those metastable states may be reached by thermal excitation if
the energy barrier between them and the ground state is comparable to or
smaller than $k_{B}T$. The saddle points between those metastable states
were investigated in an earlier paper \cite{kong} for $N\leqslant 20$. In
Fig.~\ref{Fig3} the energy and the total number of defects of different
metastable states are shown for $N=300.$ The results for the different
metastable configurations are ordered with increasing energy. Note from Fig. 
\ref{Fig3}(b) that on average the total number of defects increases with
energy, but it shows strong local variations. Only an even number of defects
are obtained, because the net topological charge is always six, and the
dipole defects (i.e. one dislocation with `--1' and `+1' defects) always
appear in pairs. Also the hexagonal position of the defects disappears (see
Fig.~\ref{Fig3}(1)) and more free dislocations are found. These defects move
from the transition region to the border (see Fig.~\ref{Fig3}(2)) or to the
central region (see Fig.~\ref{Fig3}(3)). For configurations with higher
energy, the defects arrange themselves in long chains, i.e. dislocation
lines. On average the configurations with defects at the border have a lower
energy than those with defects in the center.

We also investigated whether or not it is possible to have a configuration
with only six 5-fold disclinations and no other defects (like for example
for the $N=85$ configuration with 24 particles at the outermost ring \cite
{koulakov}). Therefore, we started our MC procedure with a perfect hexagonal
structure without any defect and then allowed it to relax to its energy
minimum. We did this for $N=281$ up to $295$ particles, because we noted
that for these particle numbers the configuration has about 42 particles in
the outer ring, which is a multiple of six, i.e. the net topological charge.
Only in such a case one can have the situation in which just six 5-fold
disclinations are present. We found that our result (from $N=281$ up to $295$
particles) never converges to a configuration with only six 5-fold
disclinations. However, this procedure indeed favorably relaxes to
configurations with 42 particles in the outer ring, often resulting in a
configuration which has less total number of defects than the corresponding
groundstate.

\section{The eigenmode spectrum}

The effect of the geometry induced defects on the eigenfrequencies (i.e. the
eigenmode spectrum) were also investigated for these large clusters. In this
system, it is well known that there are three eigenfrequencies which are
independent of $N$ \cite{26}: $\omega =0,\sqrt{2}$ and $\sqrt{6}$, which
correspond to the rotation of the system as a whole, the center of mass mode
and the breathing mode, respectively. The above modes were recently observed
experimentally \cite{Melzer}. The smallest frequency no longer correspond to
intershell rotation as in small clusters \cite{26} but to the excitation of
a vortex/antivortex pair, of which a typical mode is shown in Fig.~\ref{Fig4}%
(a). Slightly larger excitation energies may consist of multiples of such
pairs (see Fig.~\ref{Fig4}(b)). Modes with higher eigenfrequencies often
show a hexagonal structure similar to the ordering of the defects. The
motion can be concentrated around (see Fig.~\ref{Fig4}(c)) or between the
defects (see Fig.~\ref{Fig4}(d)). The local modes can be found at the six
corners of the hexagon where the defects are exactly situated (see Fig.~\ref
{Fig4}(e)). The modes in which the inner particles have larger amplitudes
than the outer particles have the largest eigenfrequencies (see Fig.~\ref
{Fig4}(f)).

The lowest eigenfrequencies of the excitation spectrum corresponding to the
ground-state configuration of the system is shown in Fig. \ref{Fig5}, as
function of the number of particles for $N$ ranging from $281$ to $307$. The
labels in Fig. \ref{Fig5} denote the total number of defects present in the
ground state. Notice that only an even number of defects are obtained as
explained before. On average, configurations with a large number of defects
have a smaller lowest eigenfrequency and are thus less stable, and vice
versa.

In this 2D lattice, all behaviors of the cluster modes can be classified as
shear-like or compression-like modes. In order to characterize the
compressional and shear parts of these eigenmodes, we calculated
respectively the divergence $\nabla \cdot \stackrel{\rightarrow }{v}$ and
the vorticity $(\nabla \times \stackrel{\rightarrow }{v})_{z}$ of the
velocity field. To calculate the velocity field, we interpolated the
displacements of Fig. 4 on a $100\times 100$ grid (thus neglecting the
constant eigenfrequency). The ``divergence and vorticity maps'' were then
calculated at every point of this matrix. Notice that pure shear or
compressional modes do not exist in the circular boundary of finite cluster.
Fig. 6(b) shows the vorticity and thus displays the shear part of the
eigenmode of Fig. 4(b). The two vortex/antivortex pairs are clearly seen.
The ``divergence map'' for this eigenmode is practically zero everywhere, as
there is no compressional part (Fig. 6(a)). This is not the case for the
eigenmodes (c) and (d) of Fig. 4. Figs. 6(c) and (e) show the ``divergence
maps'' for both eigenmodes, in which the compression and rarefaction can be
clearly seen. Both cases show no shear part (Figs. 6(d) and (f)). Figs. 6(c)
and 6(e) (see also the 3D plots at the bottom of Fig. 6) exhibit clearly
dipole type of compressional oscillations between (Fig. 6(c)) and at (Fig.
6(e)) the defect regions.

\section{Non-homogeneous melting}

Understanding the microscopic mechanism of melting has intrigued scientists
since the late nineteenth century. Special interest has been devoted to 2D
melting \cite{Rev. M.P}. Most works address infinite systems consisting of a
single layer. However, whether melting of a 2D crystal is a first order
transition and proceeds discontinuously or is a continuous transition in
which the crystal first transits into a hexatic phase retaining
quasi-long-range orientational order and then melts into an isotropic fluid,
is still an open question and a controversial issue.

In the present work we consider a finite 2D system where we take $N=291$ for
our numerical simulation. Here we present a calculation of the melting phase
diagram by performing MC simulations. In Ref. \cite{LinI2001} molecular
dynamics was used to investigate the melting of a cluster of particles
interacting through a logarithmic interaction. As compared to our Coulomb
interaction where the geometry induced defects are situated in the 3$^{th}$
and 4$^{th}$ outer shells (i.e. the transition region) and around the six
corners of the ``defect'' hexagon, in the logarithmic interacting system 
\cite{LinI2001} those defect are mainly situated in the outer two shells. In
Ref. \cite{LinI2001}, the number and type of defects were studied as
function of the noise (i.e. temperature). Here we will use several different
criteria such as the total energy, the radial dependent mean square
displacement, the bond-angular order factor and the angular square deviation
to characterize the melting behavior of the cluster.

There are several different criteria that can be used to find the melting
temperature. In order to determine the melting transition point, we
calculated the potential energy of the system as a function of temperature
(see Fig. \ref{Fig7}). In the crystalline state the potential energy of the
system increases almost linearly with temperature and then after the
critical temperature is reached ($T/T_{0}=0.01$ for $N=291$), it increases
more steeply as shown in Fig. \ref{Fig7}. This is a signature of melting and
is related to the unbinding of dislocation pairs. The dotted assurgent line
in Fig. 7 indicates the linear temperature dependence of the potential
energy for low temperatures before melting. In the upper inset Fig. \ref
{Fig7}(a), we plot $\Delta E$ which is the difference between the numerical
obtained energy and the linear T-behavior. After the melting point, $\Delta
E $ increases sup-linear.

The lower inset, Fig. \ref{Fig7}(b), shows the averaged number of defects as
function of temperature $T/T_{0}$. The number of defects were obtained as
follows. We considered 40 configurations for every temperature $T/T_{0}$.
Every 500 MC steps a new configuration was obtained. For all these
configurations the number of defects were counted. Finally we averaged over
the 40 configurations, which is the reason why the number of defects can be
non-integer. With increasing temperature, the system generates more and more
defects and after the melting point the defect number grows very fast.
Notice that two clear critical temperatures emerge from this figure at the
crossing points of the dotted lines, i.e. $T/T_{0}=0.01$ and $T/T_{0}=0.014$.

In Fig. 8 we plot typical particle trajectories for different temperatures
which shows that the melting of this system is very complex and
non-homogeneous. It clearly indicates that the melting starts around the six
corners of the hexagon which are exactly the defect regions. With increasing
temperature, the particles in the defect region start to move radially and
destroy order locally. With further increase of temperature the total system
completely melts and the order is destroyed.

In order to better describe the spatial dependence of the melting process in
our system, we separate the configuration into three regions as shown in
Fig.~\ref{Fig2}. Region I (dark grey colored hexagonal area) is comprised of
the defect-free hexagonal center; region II is a transition region with the
defects (light grey colored area), and region III consists of the outermost
two rings. For the case of $N=291$ particles region I consists of $91$
particles, region II\ consists of $116$ particles and region III\ of $84$
particles. We calculate for each region the mean square displacement $%
\left\langle u_{R}^{2}\right\rangle $, which was introduced in Ref. \cite{25}%
,

\begin{equation}
\left\langle u_{R}^{2}\right\rangle =\frac{1}{N}\sum_{i=1}^{N}\left\langle
(r_{i}-\left\langle r_{i}\right\rangle )^{2}\right\rangle /a^{2},
\end{equation}
with $a=2R/\sqrt{N}$ the average distance between the particles. Fig.~\ref
{Fig9} shows the $\left\langle u_{R}^{2}\right\rangle $ as a function of the
reduced temperature $T/T_{0}$ for the three different regions. At low
temperatures the particles exhibit harmonic oscillations around their $T=0$
equilibrium position, and the oscillation amplitude increases linearly and
slowly with temperature: the particles are well localized and display still
an ordered structure. This linear dependence is accentuated by the thin
straight lines in Fig. 9. Here, we already notice that the amplitude of the
local particle thermal vibrations in these different regions are different.
The amplitude is largest at the defect region and lowest in the center of
the cluster. Melting occurs when $\left\langle u_{R}^{2}\right\rangle $
increases very sharply with $T$. Because of the finite number of particles
one has rather a melting region, instead of a well-defined melting
temperature. After the melting ``point'', the particles exhibit liquid-like
behavior. Fig.~\ref{Fig9} exhibits three different melting temperatures
corresponding to the three different regions. Firstly region II, i.e. the
transition region containing the defects, starts to melt, then the outermost
two rings melt, and finally the hexagonal region melts. Following Ref. \cite
{Bedannov and Lozovik}, we can ``define'' a melting temperature at the point
where $\left\langle u_{R}^{2}\right\rangle \approx 0.10$ which results into
the melting temperatures $T_{melt}/T_{0}\simeq 0.0115,0.0125$ and $0.0138$
for the defect region, the outer rings and the center region, respectively.

In order to investigate the melting in the defect region in further detail
we consider two new small regions as shown in the inset of Fig. \ref{Fig10}.
One region is around the defect, the other doesn't contain a defect and is
situated between two defect regions. For $N=291$ each of the two regions
contains respectively eight and seven\ particles. In Fig.~\ref{Fig10}, the $%
\left\langle u_{R}^{2}\right\rangle $ of these two different regions show a
different melting temperature: the melting clearly starts first around the
defect as expected. The particle motion is strongly influenced by the
topological defects, i.e. the particles in the defect regions are less well
interlocked and have a larger diffusion constant than the undistorted
lattice regions and their thermal motions are easier to be excited \cite
{Moore}. Notice, that for the two separate regions a much sharper melting
behavior is found than for the intermediate region as a whole (see Fig. \ref
{Fig9}). The reason of course is that in Fig. \ref{Fig9} one averages over
defect and defect-free regions. The criterion $\left\langle
u_{R}^{2}\right\rangle \approx 0.10$ results into $T_{melt}/T_{0}\simeq
0.0118,0.0138$ for the defect and the defect-free regions, respectively.
These two melting temperatures are very similar to the melting temperature
of the transition region and the hexagonal region of Fig. 9.

The third independent parameter is the bond-orientational correlation
function. This quantity determines the type of melting transition and the
melting point for an infinite system. Our finite system is too small in
order to have a reliable analysis of the asymptotic decay of the density
correlation function. Therefore, we calculate the bond-angular order factor
which was originally presented in Ref. \cite{Nelson}, but following Ref. 
\cite{SchPRL82(1999)} we modified it into,

\begin{equation}
G_{6}=\left\langle \frac{1}{N}\sum_{j=1}^{N}\frac{1}{N_{nb}}\exp
(iN_{nb}\theta _{j,n})\right\rangle ,
\end{equation}
This quantity is calculated only for region I which exhibits a hexagon
structure, where $j$ means the $N_{nb}$ nearest neighbors of particle $i$,
for ideal hexagonal lattice $N_{nb}=6$, where $\theta _{j,n}$ is the angle
between some fixed axis and the vector which connects the $j^{th}$ particle
and its nearest $n^{th}$ neighbor.

For a perfect hexagonal system $G_{6}=1.$ In our system for $N=291$, the
initial value of $G_{6}$ is 0.96, which means that the structure in region
I\ is almost perfect hexagonal. Our numerical results (see open dots in Fig. 
\ref{Fig9}) show that $G_{6\text{ }}$ decreases linearly with increasing
temperature. When $G_{6}$ is around $0.6$, it more rapidly drops to zero
with increasing temperature. $G_{6}$ should be zero for the liquid state.
This can be compared with the infinite system where a universal melting
criterion was found in Ref. \cite{SchPRL82(1999)}: melting occurs when the
bond-angle correlation factor becomes $G_{\theta }\approx $ $0.45$, which
was found to be independent of the functional form of the interparticle
interaction. For our system the value $G_{\theta }\approx $ $0.45$ is
probably not correct because in our finite system $G_{6}$ does not drop to
zero at $T_{melt}$, but is smeared out around $T_{melt}$. Therefore, the
midpoint $G_{6}\approx 0.45/2\approx 0.225$ is expected to describe better
the melting temperature. This leads to $T_{melt}/T_{0}\simeq 0.0136$ which
is similar to the result $T_{melt}/T_{0}\simeq 0.0138$ obtained from the
radial displacement criterion.

In contrast to bulk systems, the melting scenario of small laterally
confined 2D systems was found earlier \cite{25} to be a two step process.
Upon increasing the temperature, first intershell rotation becomes possible
where orientational order between adjacent shells is lost while retaining
their internal order and the shell structure. At even higher temperatures,
the growth of thermal fluctuations leads to radial diffusion between the
shells, which finally destroys the positional order. To characterize the
relative angular intrashell and the relative angular intershell, we use the
functions as defined in Ref. \cite{25}. The relative angular intrashell
square deviation

\begin{equation}
\left\langle u_{a1}^{2}\right\rangle =\frac{1}{N_{R}}\sum_{i=1}^{N_{R}}\left[
\left\langle \left( \varphi _{i}-\varphi _{i1}\right) ^{2}\right\rangle
-\left\langle \varphi _{i}-\varphi _{i1}\right\rangle ^{2}\right] /\varphi
_{0}^{2},
\end{equation}
and the relative angular intershell square deviation

\begin{equation}
\left\langle u_{a2}^{2}\right\rangle =\frac{1}{N_{R}}\sum_{i=1}^{N_{R}}\left[
\left\langle \left( \varphi _{i}-\varphi _{i2}\right) ^{2}\right\rangle
-\left\langle \varphi _{i}-\varphi _{i2}\right\rangle ^{2}\right] /\varphi
_{0}^{2},
\end{equation}
where $i_{1}$ indicates the nearest particle from the same shell, while $%
i_{2}$ refers to the nearest-neighbor shell, $\varphi _{0}=2\pi /N_{R},$
where the number in the outermost two rings $N_{R\text{ }}$is the same and
equals 42 for our $N=291$ system. Only the two outermost rings have a clear
shell structure. Both two outer rings are strongly interlocked which is a
consequence of the 1D Wigner lattice arrangement of the two rings. From the
inset of Fig. \ref{Fig11}, one can see that the inner ring will melt before
the outermost ring. We find that the result for $\left\langle
u_{a1}^{2}\right\rangle $ of the inner ring is almost the same as $%
\left\langle u_{a2}^{2}\right\rangle $ which is the relative angular
intershell square deviation. It means that when the inner ring loses its
order, the relative order is lost simultaneously. The outermost ring can
still keep its order and it will melt at even higher temperature. Comparing
this with Fig. 9, we see that the radial and angular displacements start to
increase rapidly at approximately the same temperature. Thus for large
clusters intershell rotation will not occur below the melting temperature,
but appears at the same temperature when the radial displacements increase.

\section{Conclusion}

The configurational and melting properties of large two-dimensional clusters
of charged classical particles interacting with each other via the Coulomb
potential were investigated through the Monte Carlo simulation technique.
For the ground state configuration, a hexagonal Wigner lattice is formed in
the central area while at the border of the cluster the particles are
arranged in rings. In the transition region between them defects appear as
groups of dislocations and disclinations at the six corners of the
hexagonal-shaped inner domain. Many different arrangements and types of
defects are possible as metastable configurations with a slightly higher
energy. The particles motion is found to be strongly related to the local
topological structure. Our results clearly show that the melting of the
clusters starts near the geometry induced defects, and that three melting
temperatures can be obtained: $T_{melt}/T_{0}\simeq 0.0115,0.0125$ and $%
0.0138$ for the defect region, the outer rings and the center region,
respectively. These values are for the $N=291$ cluster. Taking a different
value for $N$ does not lead to any qualitative differences, it only
influences slightly the values for the three melting temperatures.

\section{Acknowledgments}

This work was supported by the Flemish Science Foundation (FWO-Vl), the
Belgian Inter-University Attraction Poles (IUAP-V), the ``Onderzoeksraad van
de Universiteit Antwerpen''(GOA), and the EU Research Training Network on
``Surface Electrons on Mesoscopic Structures''. We are very grateful to Dr.
I. V. Schweigert for helpful discussions. Stimulating discussions with Prof.
A. Matulis and M. Milo\v{s}evi\'{c} are gratefully acknowledged. B. Partoens
is a post-doctoral researcher of the Flemish Science Foundation (FWO-Vl).

{\bf Figure captions} 
\begin{figure}[tbp]
\caption{The ground state configurations for N=291, 300, 400, 500 particles.
The Voronoi structure is shown and the defects (i.e. disclinations) are
indicated by `+' for a 7-fold and by `-' for a 5-fold coordination number. }
\label{Fig1}
\end{figure}
\begin{figure}[tbp]
\caption{The ground state configuration for $N=291$. The dots give the
postion of the particles. Three regions are found: I (dark grey colored
hexagonal area) is comprised of the defect-free hexagonal lattice; II is a
transition region with the defects (light grey colored area), and III
consists of the outermost two rings. Obviously, there are three rings at the
border, inner region forms the hexagonal lattice, and the defects are
situated in the transition region, and they are at the exact six corners of
the hexagon. The `+1' and `-1' topological defects are represented by the
open squares and triangles, repectively. }
\label{Fig2}
\end{figure}
\begin{figure}[tbp]
\caption{The energy $E/E_{0}$ (a) and the total number of defects (b) of
different metastable states are shown for $N=300$. Three typical defect
configurations with different energy are shown at the right side of the
figure.}
\label{Fig3}
\end{figure}
\begin{figure}[tbp]
\caption{Vector plot of the eigenvectors for the cluster with $N=291$
particles for six different values of the mode number $K$. }
\label{Fig4}
\end{figure}

\begin{figure}[tbp]
\caption{Excitation spectrum of normal modes as a function of the number of
particles in the cluster. The frequency is in units of $\protect\omega 
{\acute{}}%
=\protect\omega _{0}/2^{1/2}.$ The numbers in the figure indicate the number
of defects found in the ground state of the different clusters. }
\label{Fig5}
\end{figure}
\begin{figure}[tbp]
\caption{Gray-scale contour maps of the vorticity $(\protect\nabla \times 
\stackrel{\rightarrow }{v})_{z}$ and the divergence $\protect\nabla \cdot 
\stackrel{\rightarrow }{v}$ of the velocity field of three different
eigenmodes. A corresponding 3D plot is shown for those maps which exhibit a
clear structure. }
\label{Fig6}
\end{figure}
\begin{figure}[tbp]
\caption{The potential energy($E/E_{0}$) of the 2D Coulomb cluster as a
function of temperature $T/T_{0}$ for $N=291$. The insets show $\triangle
E=E-E_{line}$ (a), and the number of defects (b) as a function of
temperature $T/T_{0}$.}
\label{Fig7}
\end{figure}
\begin{figure}[tbp]
\caption{Typical snapshots of particle trajectories for different
temperatures $T/T_{0}$ for $N=291$. }
\label{Fig8}
\end{figure}

\begin{figure}[tbp]
\caption{The mean square displacements as function of the temperature $%
T/T_{0}\ \ $for the three regions defined in Fig. 2. The open symbols are
the results for the correlation function $G_{6}$ refered to the right scale,
for the inner hexagonal region. The linear dependence at low temperature is
accentuated by the thin straight lines. The dotted curves are guides to the
eye. }
\label{Fig9}
\end{figure}

\begin{figure}[tbp]
\caption{The mean square displacements as function of the temperature $%
T/T_{0}$ for the small defect-free (open symbols) and defect regions (solid
symbols) in the intermediate region as indicated by the circular areas in
the inset. The thin straight lines show the low temperature linear
dependence. The dotted curves are guides to the eye. }
\label{Fig10}
\end{figure}

\begin{figure}[tbp]
\caption{The relative angular intrashell square deviation $\left\langle
u_{a1}^{2}\right\rangle $ and relative intershell square deviation $%
\left\langle u_{a2}^{2}\right\rangle $ of the outermost two rings as a
function of temperature for $N=291$. }
\label{Fig11}
\end{figure}

\end{document}